# Comet C/2011 J2 (Linear)

<u>Fragmentation and physical properties of the two nuclei</u>

*T. Scarmato[1]*

[1]Toni Scarmato's  Observatory, via G. Garibaldi 46,  89817 San Costantino di Briatico, Calabria, Italy

(toniscarmato@hotmail.it; Cellphone:+ 393479167369; Tel/Fax:+ 390963392102)

Version 2014 October 1

## Abstract

Comet C/2011 J2 (Linear), was discovered to the Catalina Sky Survey Observatory, based on University of Arizona. Is an hyperbolic comet passed to the perihelion at T 2013 Dec. 25.4020 TT.  The distance q from the Sun is 3.443732 A.U., that mean that don't never cross the "snow line". Also, the Sun's gravitational force is not such to determine a strong stress on the cometary nucleus.  A possible fragmentation at that distance from the Sun in general, is unlikely, but on 2014 September 19[th], the CBAT 3979 by D. Green of the ICQ, report the discoveries of a fragment companion of the comet's main body. In the following days others observers confirms the presence of the secondary "body". (F. Manzini et al. September 2014, CBET 3986, 2014 September 24th)

In the observation of the comet on 2014 September 28, using a 25 cm Newton and CCD with R photometric filter, I detect the secondary "body" and assuming as possible scenario the presence of a solid fragment in the cloud of material of the fragmentation, I measured some physical parameters, the magnitude in R band and Af(rho) value of the two "nuclei". Here are presented the preliminaries data.

## 1) Introduction

In the night on 2014 September 28[th], using the following setup, I observed the comet coming from the Oort cloud C/2011 J2 (Linear).

### Basic setup used for observations

$$FOV(') = \frac{3428 * \dim(mm)}{focal(mm)}, \; scale(arc\sec/pixel) = \frac{FOV}{\dim(pixels)} * 60$$

### Toni Scarmato's Observatory (T. Scarmato Observer)

| Parameters ATIK 16IC mono | |
|---|---|
| Area in Pixel array | 659X494 (325,546 pixels squares) |
| Pixel size | 7.4x7.4 micron |
| Full well depth | 40.000e |
| Dark current | <1e per second at −25°C |
| Peak spectral response | 500 nM |
| Quantum efficiency | >50% a 500 nM |
| A−D converter | 16 bits |
| Readout noise | 7e |
| Anti−blooming | yes |
| Cooling | yes |
| CCD Type | Sony ICX−424AL |
| CCD size (dim area sensitive) | 4,8x3,7 mm |

| Parameters Telescope | | Parameters Filter Rc | |
|---|---|---|---|
| Aperture | 250 mm | Productor | Schuler |
| Focal Lenght | 1200 f/4.8 | Band | Large |
| Scale | 1.27 arcsec/pixel | Lambda peak | 5978 A |
| Optic | Newton | FWHM | 1297 A |
| Type | Reflector | | |
| FOV (Field of View) | 14'x11' | | |

After the CBAT 3979 publication I planned an observation of the comet based on the predicted magnitude and position on the sky. The 2014 September 28 night the comet was high in the sky about 75 deg in Andromeda constellation. The sky was clear and transparent with a PSF of about 1.5 arcsec. The magnitude limit on the stacked image was about 20 in R band (See fig. 1). Only 3 images of the series were excluded due to the bad tracking of the telescope. The single exposure was of 240 sec, so we have 7 good images calibrated with darks and flats and stacked using Astrometrica tool to obtained a very good alignment with subpixels precision.

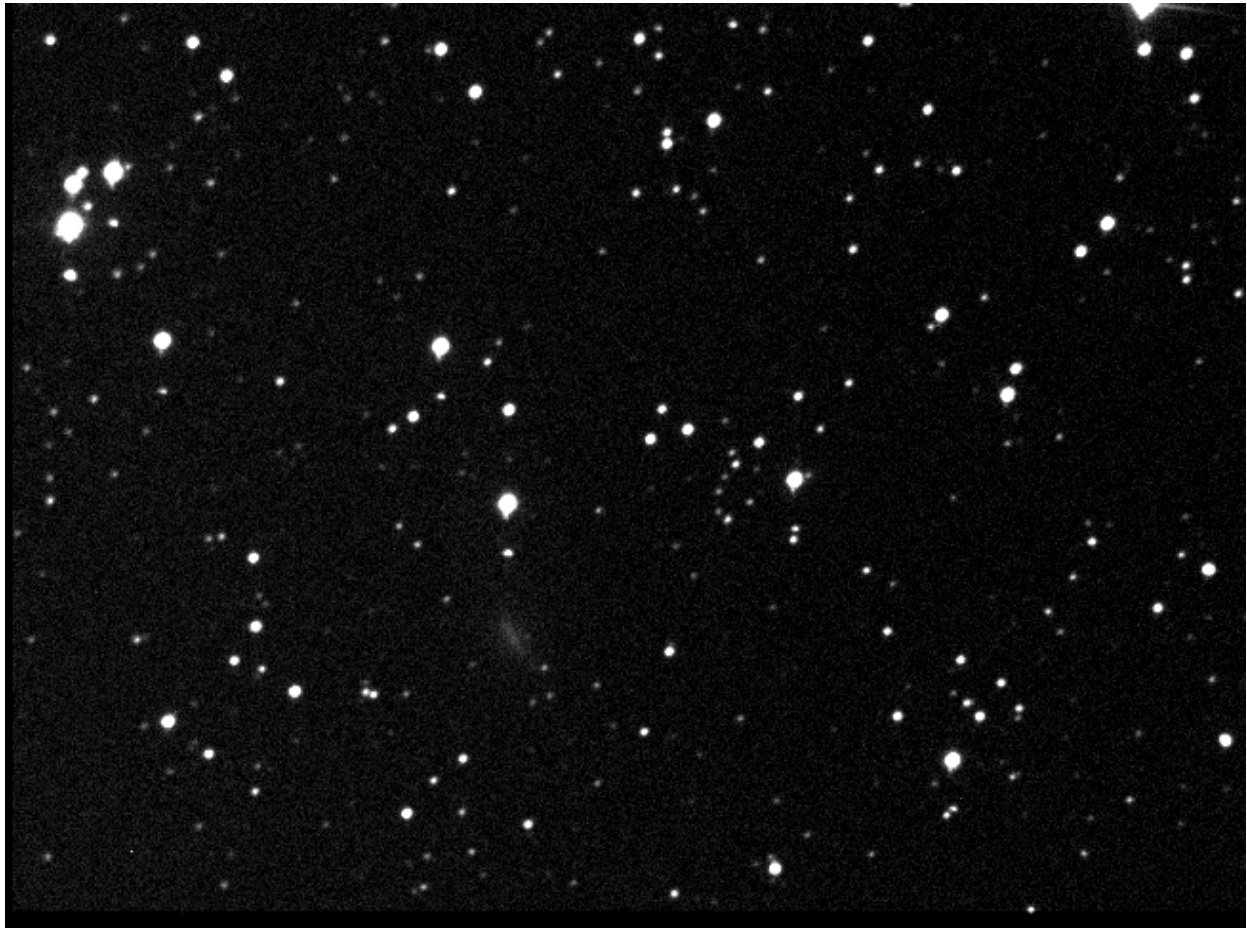

Fig. 1 − Original image obtained stacking 7x240 sec fits images aligned on the stars. Star magnitude limit on the image is about 20 in R band. Is visible the comet track. North is up and Est is to left.

## 2) Formulae and algorithms

Observation started on 2014 September 28 at 20:31:31 UT and ended to 21:08:14 UT. In total were taken 10 images of which 3 excluded. To calibrate the images were taken 13 flats and 7 darks images. No bias need. The images will then be calibrated with the masterdark, masterflat The temperature varied during the observation of about 0.2 °C. Time exposure of 240 sec to obtained the better S/N ratio.

The comet was at 3.4924 A.U. from the Earth and to 4.2636 A.U. from the Sun and the phase angle was 9.5 deg. The star used is USNOB1/Tycho2 1343-0518118/3229-01009-1,Rmag=11.421+/-0.037 Bmag=12.652+/-0.042 in the same field of the comet. Using the formula

$$Flux = \frac{AduComet}{\sum Adu(ref)} \quad \text{(Eq. 1)}$$

and **magcomet = - 2.5 * LOG (FLUX) + magref** (where magref is the magnitude of the reference star in R band,, we have the magnitude of the comet. To compute the error, I used the standard deviation at 1 sigma of precision.

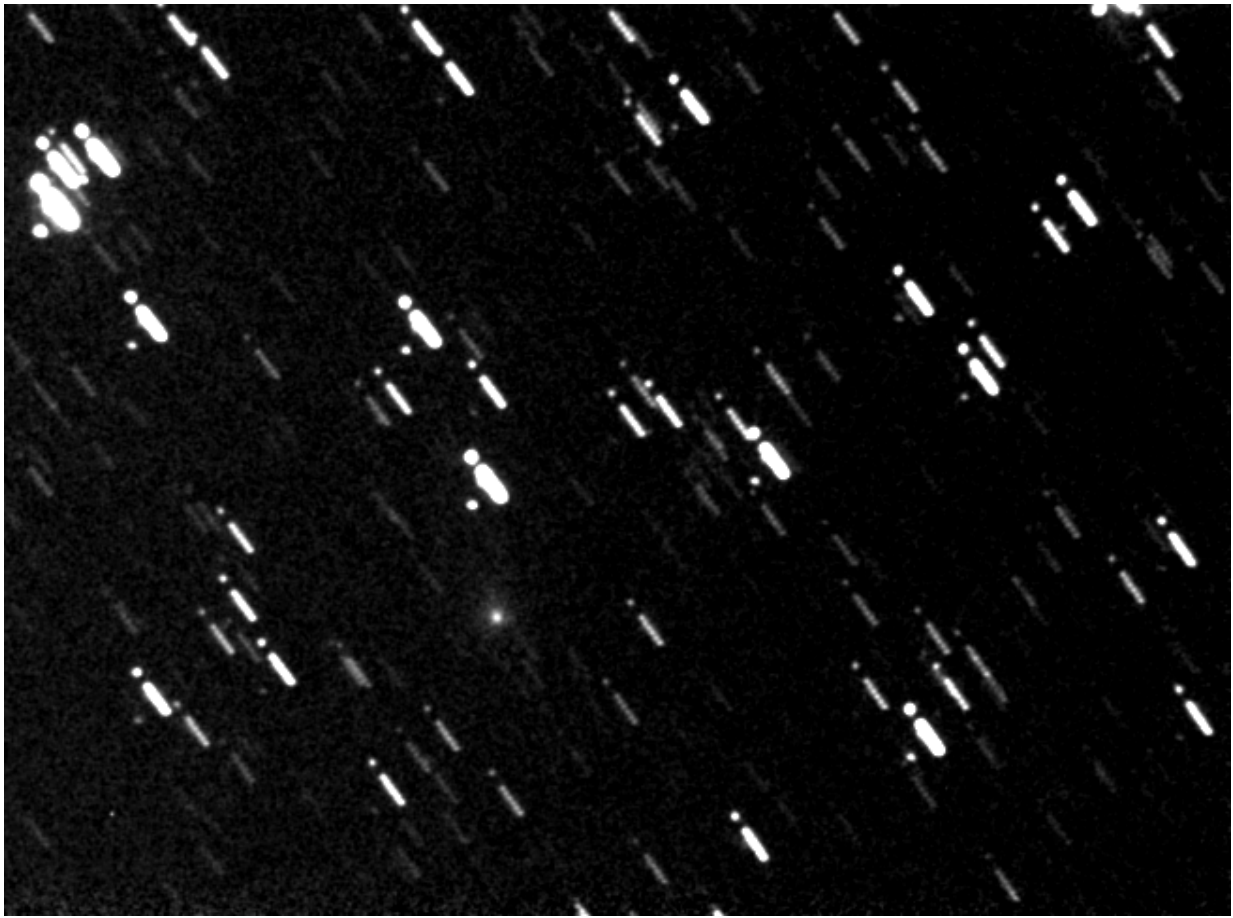

Fig. 2 – Stacked image with Astrometrica tool, 7x240 sec aligned on the comet center. North is up and Est is to left.

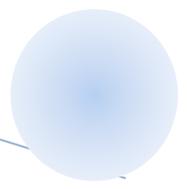

To compute the **Af(rho)** value (**A'Hearn 1984**) was used the following formula:

$$Af(rho) = (2.467 x 10^{19}) x \frac{Rs^2 x D}{ap} x \frac{FluxComet}{FluxSun} \quad \text{(Eq. 2)}$$

that with simple transformation become,

$$Af(rho) = (2.467 x 10^{19}) x \frac{Rs^2 x D}{ap} x 10^{0.4 x (Ms-Mc)} \quad \text{(Eq. 3)}$$

where Rs is the Sun−Comet distance in A.U., D the Earth−Comet distance in A.U, **ap** is the used aperture in arcsec, **Ms=−27.09** the magnitude in R band of the Sun and **Mc** the measured magnitude in R band of the comet. The method to measure the nucleus of a comet also with amateur images at lower resolution, was discussed in **Toni Scarmato 2014**, http://arxiv.org/abs/1405.3112. Measure the ADU for the nucleus and using the star in the FOV of the images and his R magnitude compute the radius of the comet. The equations are;

$$R_n = \frac{1.496 x 10^{11}}{\sqrt{p}} x 10^{0.2 x (M_s - H)} \quad \text{(Eq. 4)}$$

where

$$H = m - 5 x \log(R_s) x D - \alpha\beta \quad \text{(Eq. 5)}$$

is the absolute magnitude of the nucleus in R band, and

$$\alpha\beta = -2.5 x \log[\Phi(\alpha)] \quad \text{(Eq. 6)}$$

where $\Phi(\alpha)$ is the phase function, $\alpha$ is the phase angle, **m** is the apparent magnitude of the nucleus measured from the observations in R band and **β** the phase coefficient. We assumed $R_s$ =1 A.U., β=0.04 mag/deg and the albedo **p=0.04** as standard values.

Also is possible to use a more simple formula (I. Ferrin " "The location of Oort Cloud Comets C/2011 L4 (Panstarrs) and C/2012 S1 (ISON), on a Comets´ Evolutionary Diagram");

$$R_n = \frac{\sqrt{10^{7.654-0.4xH}}}{2} \quad \text{(Eq.7)}$$

## 3) Image enhancement; 1/rho model background subtraction, Bicubic Interpolation, Convolution and PSF

After having calibrated the original fits images with dark, flat and bias, we applied an algorithm that extracts the pixels value of the coma, subtracts the background value and multiplies for the cometcenter distance to create a new image with the computed values. After this, we used a crop of 40x40 pixels image centered on the nucleus position, to apply the Bicubic Interpolation, Convolution and PSF.

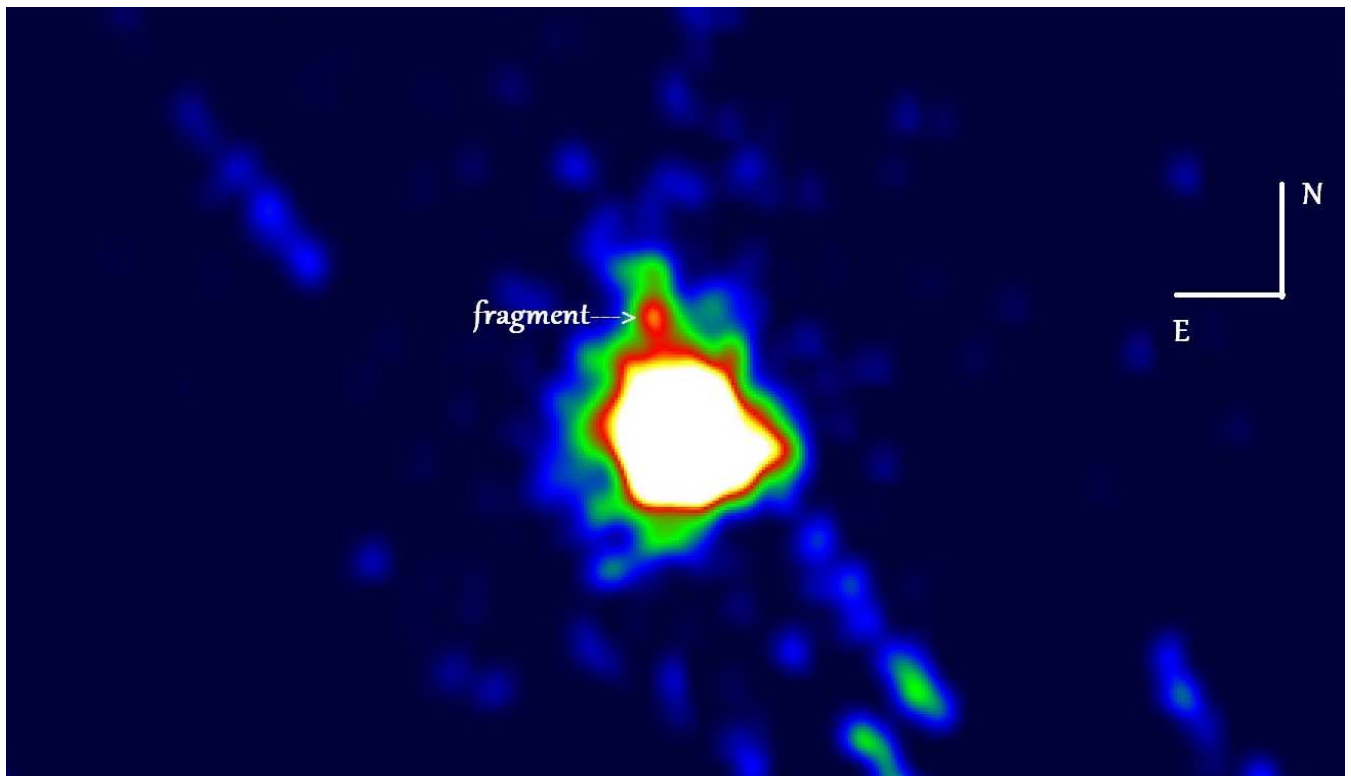

Fig. 3 – Cropped image that show the main nucleus and the fragment with a clear central condensation. Comet position predicted AR= 23h 16m 33.85s, decl.= +44° 19' 08.11"

This procedure permits the construction of a finer structure of the coma and to define the values of the sub-pixels around the cometcenter pixel that contain the nucleus. In this manner, we can obtain details of the coma structure and the residuals between the brighter pixel containing the nucleus and the other pixels around containing the nucleus contribution to the brightness of the coma.

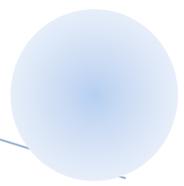

# 4) Discussion of the results

The first step, as mentioned above, was to calibrate and determine the astrometry and photometry in the original fits images stacking all the well resolved and tracked images, using Astrometrica tool. The fig.4 show a magnification of the original image in which is possible to identify the brighter pixel that contain the nucleus and the fainter fragment pointed out from the arrow.

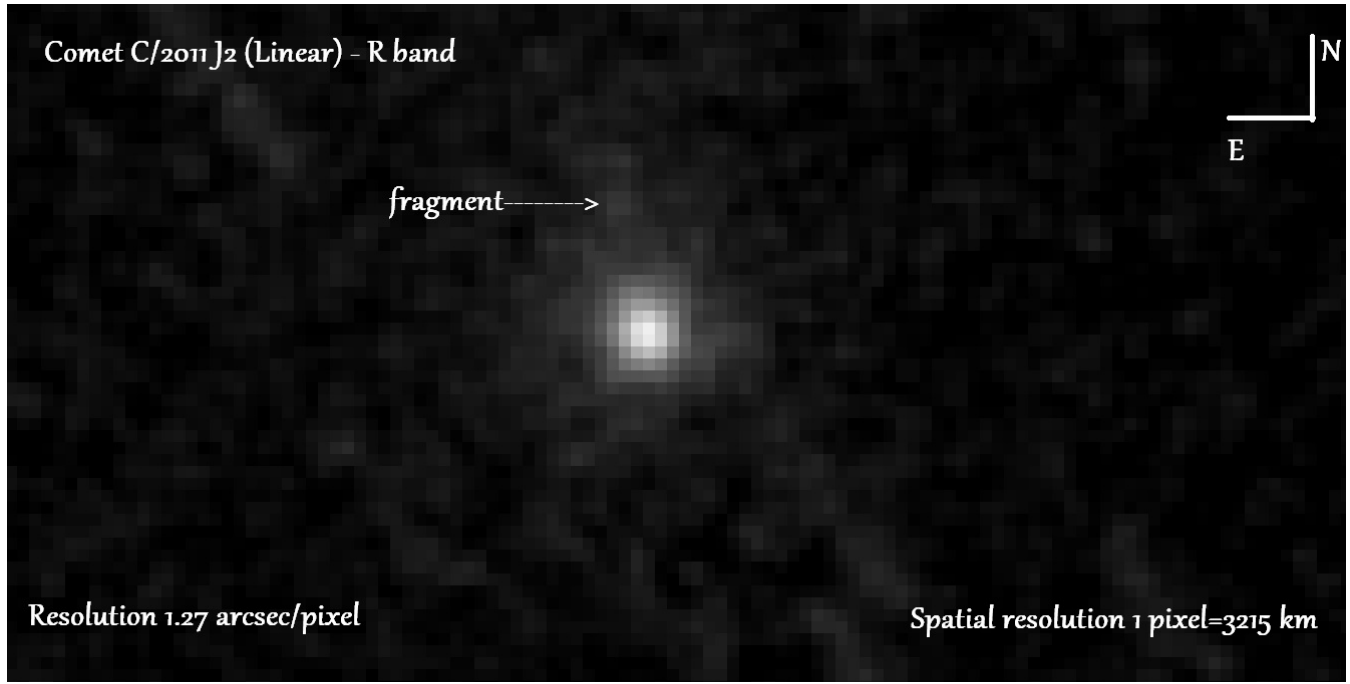

Fig. 4 – Magnification of the original fits 7x240 sec image in focus stacked with Astrometrica tool, calibrated also astrometrically.

The brighter central pixel of the main component was located at the position

<div style="text-align:center">AR=23 16 18.981 decl.=+ 44 21 20.78</div>

and the centre of the fragment at the position

<div style="text-align:center">AR=23 16 18.534 decl.=+ 44 21 08.36.</div>

So the offsets of the fragment are respectvely in AR=0.447 arcsec and in decl.=12.42 arcsec. Using the photometry differential of aperture, as discussed in Toni Scarmato 2014, http://arxiv.org/abs/1409.2693, for ap=7 pixels, equal to 8.9 arcsec, spatial extension of 22500 km of coma, we have the following results;

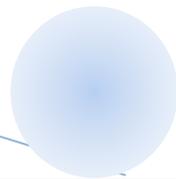

| Comet C/2011 J2 (Linear) and Fragment – Photometry differential of aperture – R band | | | |
|---|---|---|---|
| 2014 September 28 – JD 2456929.35523 | | | |
| Aperture 7pixels=8.9arcsec=22500km | | | |
| Reference star | | | |
| USNOB1/Tycho2 1343-0518118/3229-01009-1, Rmag=11.421+/-0.037 Bmag=12.652+/-0.042 | | | |
| Component | Magnitude | Afrho cm (dust production) | Qdust (kg/day) |
| C/2011 J2 | 13.504 +/- 0.021 | 10191 +/- 530 | 8.8*10^8 |
| Fragment | 15.015 +/- 0.045 | 2535 +/- 120 | 2.2*10^8 |

Tab. 1 – Photometry differential of aperture and Af(rho) computation based on the formulae discussed above in section 2.

The table above show the strong activity of the two bodies regarding the dust production. Gived the low phase angle value of only 9.5 deg, we can assume that the light coming from the Sun is completely reflected from the system nuclei/coma toward the Earth direction. Another night of observation will able me to compute the relative speed between the two "comets". For this look interesting the profile shown in the next figure.

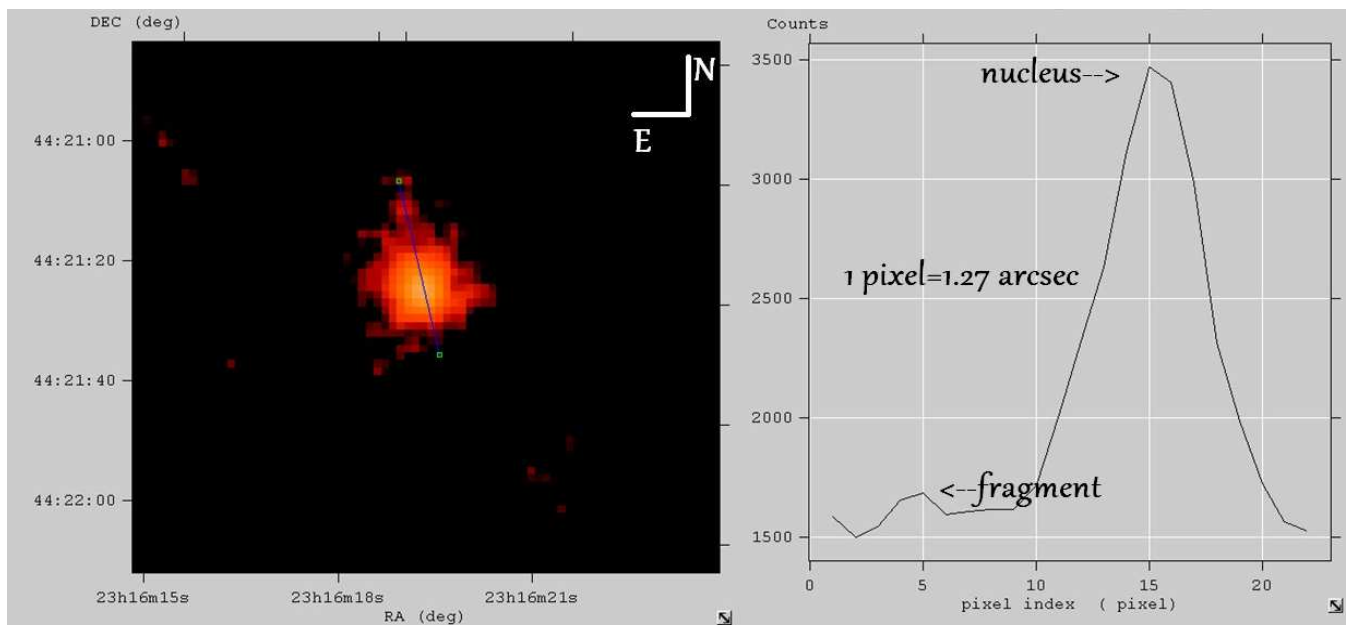

Fig. 5 – Astrometric position of the main nucleus and fragment with the profile and the distance in pixels of the two peaks, about 10 pixels, equal to 12.7 arcsec, in perfect agreement with the astrometric reduction of Astrometrica tool.

Assuming a scenario that consider the secondary condensation as a solid body, using the algorithms described in Toni Scarmato 2014, "Sungrazer Comet C/2012 S1 (ISON): Curve of light, nucleus size, rotation and peculiar structures in the coma and tail" http://arxiv.org/abs/1405.3112, I was able to determine the size of the "nuclei". (See Fig.6)

| COMET C/2011 J2 (Linear) - 2014 September 28 - Observer T. Scarmato | | | | | | | | Mean Nucleus | | |
|---|---|---|---|---|---|---|---|---|---|---|
| Table 1 | ADU | Band R | albedo p=0,04 | | 1AU=149,577,000 km | | | | | |
| axis | residual | magapp | H (absolute) | Rcomet (m) | err +/- | rho^a | Phase ° | D (AU) | r (AU) | |
| x | 789 | 18,197 | 17.817 | 781 | 138 | a=-1 | 9.5 | 3,4924 | 4,2636 | |
| y | 591 | 18,523 | 18.143 | 672 | 119 | a=-1 | 9.5 | 3,4924 | 4,2636 | |
| xy | 1270 | 17,692 | 17.312 | 985 | 174 | a=-1 | 9.5 | 3,4924 | 4,2636 | |
| yx | 1479 | 17,527 | 17.147 | 1063 | 188 | a=-1 | 9.5 | 3,4924 | 4,2636 | |
| Average | 1032 | 17,985 | 17.605 | 875 | 154 | a=-1 | 9.5 | 3,4924 | 4,2636 | |
| COMET C/2011 J2 (Linear) - 2014 September 28 - Observer T. Scarmato | | | | | | | | Fragment | | |
| Table 1 | ADU | Band R | albedo p=0,04 | | 1AU=149,577,000 km | | | | | |
| axis | residual | magapp | H (absolute) | Rfrag (m) | err +/- | rho^a | Phase ° | D (AU) | r (AU) | |
| x | 251 | 19,453 | 19.073 | 438 | 77 | a=-1 | 9.5 | 3,4924 | 4,2636 | |
| y | 84 | 20,641 | 20.261 | 253 | 45 | a=-1 | 9.5 | 3,4924 | 4,2636 | |
| xy | 295 | 19,277 | 18.897 | 475 | 84 | a=-1 | 9.5 | 3,4924 | 4,2636 | |
| yx | 318 | 19,196 | 18.816 | 493 | 87 | a=-1 | 9.5 | 3,4924 | 4,2636 | |
| Average | 237 | 19,642 | 19.262 | 415 | 73 | a=-1 | 9.5 | 3,4924 | 4,2636 | |

Fig.6 – The table show the apparent and absolute magnitude of C/2011 J2 main body and the fragment, the size of the nucleus radius, assumig an albedo p=0.04.

How it is possible see in the table above, the comet was at r=4.2636 A.U from the Sun, largest than the "snow line" distance. A fragmentation at that distance is unlikely also if in the last times several comets showed not only a strong activity but also signs of fragmentation at r>6 A.U. (See C/2012 S1 ISON comet). Looking at the size of the two "comets", the average dimensions of the original body are about 2.5 km in diameter. Not a big comet. Other observations and further analysis of the data will give us more information about the strong event happened on C/2011 J2 (Linear) comet.

___________________________________